\documentclass[conference]{IEEEtran}
\IEEEoverridecommandlockouts
\usepackage{cite}
\usepackage{amsmath,amssymb,amsfonts}
\usepackage{algo}
\usepackage{subcaption}
\usepackage{float}

\def\argmin{\operatornamewithlimits{arg\,min}}
\usepackage{graphicx}
\usepackage{textcomp}
\usepackage{hyperref}
\usepackage{xcolor}
\usepackage{braket}
\def\BibTeX{{\rm B\kern-.05em{\sc i\kern-.025em b}\kern-.08em
    T\kern-.1667em\lower.7ex\hbox{E}\kern-.125emX}}
\begin{document}

\title{Post-Disaster Repair Crew Assignment Optimization Using Minimum Latency\\
\thanks{This material is based upon work supported by the Discovery Partners Institute (DPI) Science Team Seed Grant Program. DPI is part of the University of Illinois System.}
}


\author{\IEEEauthorblockN{Anakin Dey}
\IEEEauthorblockA{\textit{University of Illinois Urbana-Champaign}\\
Urbana, USA \\
anakind2@illinois.edu}
\and
\IEEEauthorblockN{Melkior Ornik}
\IEEEauthorblockA{\textit{University of Illinois Urbana-Champaign}\\
Urbana, USA \\
mornik@illinois.edu}
}

\maketitle

\begin{abstract}\boldmath 
Across infrastructure domains, physical damage caused by storms and other weather events often requires costly and time-sensitive repairs to restore services as quickly as possible. While recent studies have used agent-based models to estimate the cost of repairs, the implemented strategies for assignment of repair crews to different locations are generally human-driven or based on simple rules. In order to find performant strategies, we continue with an agent-based model, but approach this problem as a combinational optimization problem known as the Minimum Weighted Latency Problem for multiple repair crews. We apply a partitioning algorithm that balances the assignment of targets amongst all the crews using two different heuristics that optimize either the importance of repair locations or the travel time between them. We benchmark our algorithm on both randomly generated graphs as well as data derived from a real-world urban environment, and show that our algorithm delivers significantly better assignments than existing methods.
\end{abstract}


\section{Introduction}

Extreme weather events often result in significant damage over a large area \cite{extreme-events-happen, climate-change}. In order to minimize costs and the loss of life, the repair of infrastructure such as road, power, and water networks is a central task. The importance of this task is made even greater due to the increasing frequency and intensity of these events correlating to climate change \cite{climate-change}. Following a significant adverse event, emergency managers must dispatch a limited number of crews to efficiently repair all damaged locations \cite{estimation}, noting that the locations may have differing importance, e.g., because of differing population densities. This paper tackles the resulting problem of \textit{crew assignment}: our goal is to automatically determine a strategy for assigning repair crews to locations in need of repair work in order to minimize the effects of the sustained damage.

Repair crew routing in damage restoration has been the subject of significant prior research. A popular approach \cite{linear-program-positions, linear-program-networks, linear-program-stations} has exploited linear programming, reducing the problem to a system of equations describing the constraints of the problem. The equations describe the repair time for various locations, the number of repair crews starting from a given position, and indicator variables representing whether a particular location has been repaired. However, they do not consider the case where targets may be of various levels of importance.

In this paper, we consider an \textit{agent-based model (ABM)} to solve the problem of repair crew assignment. Doing so allows us to place different importance on different repair locations and directly study the sequences of locations that are repaired, thus uncovering the \textit{dynamics of the overall repair process over time}. Our work builds on previous ABMs \cite{estimation} and \cite{prediction}. However, these prior ABMs largely dealt with simulating the effects of several simple repair assignment strategies. Instead, we consider the problem of \textit{generating optimal assignments} and find  approximate solutions by relating it to a class of optimal routing problems in computer science. Our method relies on iteratively improving the partition of repair locations into crews by a \textit{transfers-and-swaps} method \cite{transfers-swaps}, followed by using a heuristic from  \cite{estimation, prediction} to generate a routing plan for each crew separately.

On a number of synthetic examples, as well as an urban environment modeled after Champaign, Illinois, USA, we show that our method yields assignments of locations for repair crews that outperform previously considered strategies in a number of metrics, including average waiting time until service restoration.

The paper is organized as follows: Section \ref{prob_work} formally describes the problem, notation, and the assumptions we make. In Section \ref{algorithm}, we describe the proposed two-stage partitioning and assignment algorithm which balances the work assigned to different crews. Finally, Section \ref{res} contains benchmarks of our algorithm against prior algorithms on both randomly generated graphs and a real-world urban environment. We conclude in Section \ref{future} with a discussion on possible extensions of this preliminary work.

\section{Problem Description and Prior Work}
\label{prob_work}

We follow the model proposed in \cite{estimation, prediction} and study the problem of repair crew assignment by modeling the repair locations (``targets'') and the movements of crews by a node-weighted and edge-weighted graph. The edge-weights represent travel time between each of the locations; in this preliminary work, we assume a constant fixed speed of each crew. We also assume that repair times are known in advance and add the repair time of each target to the weight of incoming edges. Each target has an associated node-weight representing the importance of that location. The goal is to assign each repair crew to a sequence of target locations so that the average time to repair completion, weighted by target importance such as population density or infrastructure criticality, is minimized.

\subsection{Problem Formulation}
\label{probdes}

We model our problem using a complete node-weighted and edge-weighted graph $G = (V, E)$. We denote the number of nodes by $n$; each node representing one of the repair targets. The \textit{weight of a node} $v\in V$ on the graph is $w(v)$ and the \textit{weight along edge} $e \in E$ from $u \in V$ to $v \in V$ is $d(e)$, also denoted as $d(u \to v)$. The node-weights represent the importance of a location and the edge-weights represent the sum of the shortest travel time between targets, possibly calculated using Dijkstra's algorithm on the overall area, and the repair time for the destination target. 

Consider a single repair crew traveling on a path $\phi$ denoted by a list of targets $\phi = \left[ v_1, v_2, \ldots, v_k \right]$ which starts at some designated start node $s = v_1$ representing a repair depot. For formal reasons, we assign $w(s)=0$, indicating that $s$ does not contribute to the total cost of the repair of other nodes. The \textit{latency} of a node $v_i$ along $\phi$ is denoted by $l_\phi(v_i)$ and defined as \textit{the total travel time from $v_1$ to $v_i$} along $\phi$: \begin{equation*}
    l_\phi(v_i) = \sum_{j = 1}^{i - 1} d(v_j \to v_{j + 1}).
\end{equation*} 
The notion of latency describes the waiting time of a target until it is visited and repaired by an agent. For node-weighted graphs, we define the \textit{weighted latency} of node $v_i$ with weight $w(v_i)$ along path $\phi$ by $w(v_i) l_\phi(v_i)$. In the setting of post-disaster repair, weighted latency describes the cost accrued by a target as a product of target importance and travel time to a target. The total cost of a path $\phi$, called the \textit{weighted latency over path $\phi$}, is thus defined by \begin{equation}\label{wlp}
    \textsc{wlp}(\phi) = \sum_{i = 1}^k w(v_i) l_\phi(v_i).
\end{equation} 
The value of \eqref{wlp} is a weighted total of waiting times of customers at a location before service is restored to that location. The problem of solving for a path through all nodes in a graph that minimizes \eqref{wlp} is known as the \textit{minimum weighted latency problem} (MWLP) \cite{MWLP-cases}.

Given $m$ agents operating on graph $G$, the \textit{agent partition} $\mathcal{P} = \Set{\mathcal{V}_1, \ldots, \mathcal{V}_m}$ is a set of subsets of $V$ satisfying
\begin{equation}\label{partition_def}
    \begin{aligned}
        &\bigcup_{i = 1}^m \mathcal{V}_i = V, \\
        &\mathcal{V}_i \cap \mathcal{V}_j = \{s\} & \textrm{for all } 1 \leq i < j \leq m.
    \end{aligned}
\end{equation}
We call the set of all agent partitions $\mathcal{B}$. An agent partition describes the order of targets that each agent needs to visit, starting from the joint starting location $s$. Due to the differing importance of targets, the repair strategy performance also depends on the sequence in which the targets are visited. We thus define an \textit{agent assignment} $\pi = \set{\phi_1, \ldots, \phi_m}$ where each path $\phi_i = \left[s, v_{i, 2}, \ldots, v_{i, n_i}\right]$ is an ordering of $\mathcal{V}_i$ in an agent partition $\mathcal{P} = \Set{\mathcal{V}_i, \ldots, \mathcal{V}_m}$. We naturally call $\phi_i$ an \textit{assignment on} $\mathcal{V}_i$. An assignment $\pi$ satisfies the following: 
\begin{equation}\label{assignment_def}
    \begin{aligned}
        & \pi = \set{\phi_1, \ldots, \phi_m}, \textrm{ where each } \phi_i \textrm{ is an ordered list} \\
        & \text{path } \phi_i \text{ begins with } s \textrm{ for all } 1 \leq i \leq m, \\
        &\bigcup_{i = 1}^m\bigcup_{j=2}^{n_i}{v_{i,j}}=V\setminus \{s\},\\
        & \text{each } v \in V \setminus \{s\} \text{ is in exactly one } \phi_i.
    \end{aligned}
\end{equation}
We denote the set of all possible agent assignments $\pi$ by $\Pi$. We can now proceed to define a metric of \textit{total time-cost of a repair}, indicating the sum of time accrued until each target is repaired, weighted by target importance. Given some $\pi = \set{\phi_1, \ldots, \phi_m} \in \Pi$, we define this cost by

\begin{equation}\label{wlp_sum}
    \textsc{wlp\_sum}(\pi) = \sum_{\phi_i \in \pi} \textsc{wlp}(\phi_i).
\end{equation}
Our goal is thus to solve the following problem:

\begin{equation}\label{optimization-sum}
    \argmin_{\pi \in \Pi}\left\{ \textsc{wlp\_sum}(\pi) \right\}.
\end{equation}
Because it is a multi-agent generalization of the MWLP, we call this combinational optimization problem, given $m$ repair crews, the \textit{minimum weighted latency problem for $m$-agents} ($m$-MWLP). Before continuing to discuss this problem, we recognize that the connection between the real-world problem of repair assignment and the formally stated $m$-MWLP relies on a number of assumptions. We briefly recount these assumptions.

\subsection{Model Assumptions}
\label{assumptions}

The overarching feature of the proposed model is that all relevant system information is \textit{a priori known, deterministic, and time-invariant}. Namely, we assume that the crew travel times between any two targets are known in advance and do not change, that the time needed to repair each target is known even before the crew is assigned to the target, and that the importance of each target is perfectly modeled in advance. We also assume that every repair crew can handle every type of outage. These assumptions are used in much of the prior literature for the MLP and MWLP \cite{ride-and-delivery, MLP-DP, the-MLP}, as well as outage restoration literature \cite{estimation, prediction}. 

In addition to the assumption of complete knowledge, we also assume that service at each target is independent of other targets. In other words, if an agent repairs a target, the service to the surrounding area --- which corresponds to the target importance --- is immediately restored. Such an assumption is present in \cite{estimation}. Finally, we assume that each repair crew operates continuously without breaks until all the assigned locations are repaired, and that all crews are equal in every way, i.e., that the time to repair a given target or to move from one given target to another is the same for all crews. These assumptions also follow \cite{estimation}.

We now return to the proposed $m$-MWLP model and describe previous investigations for its exact and approximate solutions.

\subsection{Recent Work}

It has been shown in \cite{MLP-NP-Complete} that the MLP is NP-complete, implying that the MWLP --- the weighted generalization of the MLP --- is also NP-complete. This result in turn implies that the $m$-MWLP --- as the multi-agent generalization of the MWLP --- is NP-complete. Hence, solving \eqref{optimization-sum} via a brute force methodology is computationally infeasible, even for small graphs. Exact polynomial-time algorithms with respect to graph size for the MLP have been found for some special cases in \cite{fixed-graph, delivery-tree, TRP-Complexity, the-MLP}, and also for the MWLP in \cite{MWLP-cases}. However, apart from improvements to exponential-time algorithms through branch-and-bound methods and dynamic programming \cite{MLP-DP, MWLP-cases}, the majority of the work has focused on polynomial-time heuristics \cite{approx-and-random,parallel-programming,paths-and-trees}. Such work is also largely focused on single-agent problems, with limited exceptions such as a case of $m$ agents on a fixed $n$-node path or cycle \cite{MWLP-cases}. Existing work on outage restoration \cite{estimation, prediction} considers multiple crews, but does not consider optimal assignments, instead only analyzing simple assignment strategies based on human intuition. 

In contrast to previous work on outage restoration, our work builds on existing efforts on optimizing graph node partitions with respect to a metric of cost for multi-agent, multi-target problems. Namely, we use the \textit{transfers-and-swaps} method of iterative partition and assignment improvement, previously applied to problems such as the \textit{Traveling Salesman Problem} for $m$ agents \cite{transfers-swaps}, as well as the problem of multi-agent, multi-target reachability in a stochastic environment \cite{transfers-swaps-MDP}. However, there has not been any work towards optimizing the $m$-MWLP in the general case. We describe our method in the following section.

\section{Proposed Algorithm}
\label{algorithm}

The model that we developed in Section \ref{probdes} allows us to find \textit{the optimal repair crew assignment}, expressed by a solution to the $m$-MWLP in \eqref{optimization-sum}, \textit{in finite time} by evaluating all possible partitions and assignments. However, as described above, the optimal solution is infeasible to compute even for smaller graphs due to the NP-completeness of the problem. In this section we develop a heuristic approach to generate partitions and assignments.

Our method draws from the transfers-and-swaps algorithm of \cite{transfers-swaps}, which optimizes a partition $\mathcal{P}$ by making incremental changes to its subsets to reduce the cost incurred by the partition. It does not consider the whole partition at a time, but attempts to reduce the maximum cost among every pair of subsets $(\mathcal{V}_i,\mathcal{V}_j)$. The algorithm uses two main operations. The first operation,\textit{transfer}, moves a node from $\mathcal{V}_i$ to $\mathcal{V}_j$. The second, \textit{swap}, takes a node in $\mathcal{V}_i$ and moves it to $\mathcal{V}_j$ and simultaneously takes a node in $\mathcal{V}_j$ and moves it to $\mathcal{V}_i$. 

The notion of cost in $m$-MWLP, defined in \eqref{wlp_sum}, depends on ordered assignments \eqref{assignment_def} whereas the transfers-and-swaps algorithm works with unordered partitions \eqref{partition_def}. Thus, we begin by developing a simply computable notion of the \textit{cost of a subset $\mathcal{V}_i$ given an assignment} $\phi_i$ on $\mathcal{V}_i$.

\subsection{Approximate Partition Subset Cost}

Unfortunately, a natural idea of identifying the cost of $\mathcal{V}_i$ with the cost of the optimal assignment on $\mathcal{V}_i$ already presents an NP-hard problem as computing such a cost amounts to solving the MWLP for a single agent. It is thus not appropriate for use in a heuristic effort such as transfers-and-swaps. Instead, we propose to use the cost incurred by intuitive strategies on a fixed partition as a proxy for minimal partition cost. 

Subsets of nodes $\mathcal{V}_i \subseteq V$ induce complete subgraphs $G_i \subseteq G$ \cite{transfers-swaps}. Given these subgraphs, we consider two natural heuristics. The first is a \textsc{nearest\_neighbor} heuristic. For each partition element $V_i$, we construct a path starting at $s$ and visit the nearest unvisited node in $G_i$ --- i.e., the node with the minimal edge-weight from the current node --- at each step, until all nodes in $G_i$ are visited. The second heuristic is a \textsc{greedy} one. It works similarly to \textsc{nearest\_neighbor} but attempts to minimize \eqref{wlp} by visiting the unvisited target with the highest importance at each step.

We note that both the \textsc{greedy} and \textsc{nearest\_neighbor} heuristics can be calculated in polynomial time with respect to $n$. Previous work on outage restoration \cite{estimation, prediction} uses these exact heuristics as possible strategies for target assignment. However, they do not partition the agents' tasks explicitly and simply consider all agents operating on all of $G$ at the same time. We describe prior strategies in more detail in Section~\ref{res}, where we will show that our method of explicit partitioning may yield significantly better results than these methods.

Applying these two heuristic strategies generates paths $\phi_i$, and thus an assignment $\pi$, for a given partition $\mathcal{P}$. The cost of partition $\mathcal{P}$ is then obtained by \eqref{wlp_sum}, with --- naturally --- a different cost depending on the heuristic strategy. With this proxy for the cost of a partition, we now describe our adaptation of the transfers-and-swaps algorithm.

\subsection{Iterative Optimization}

Our method of optimizing agent partitions begins by fixing a cost heuristic for partitions, \textsc{greedy} or \textsc{nearest\_neighbor}, and randomly choosing a starting partition $\mathcal{P}\in\mathcal{B}$.

Starting from the random partition $\mathcal{P}$, the first step performs transfers-and-swaps. Namely, for each pair $(\mathcal{V}_i,\mathcal{V}_j)$ of subsets in $\mathcal{P}$, we identify the best node to transfer from $\mathcal{V}_i$ to $\mathcal{V}_j$ and update $\mathcal{P}$ accordingly, until no further transfers yield improvements. Node $v^* \in \mathcal{V}_i$ is \textit{the best node to transfer} if its transfer maximally reduces the maximum of the cost of two subsets, i.e., if $v^*$ maximizes $$\max(c(\mathcal{V}_i),c(\mathcal{V}_j))-\max(c(\mathcal{V}_i\backslash\{v\}),c(\mathcal{V}_j\cup\{v\}))\textrm{,}$$ where $c$ is the chosen heuristic. No transfer from $\mathcal{V}_i$ to $\mathcal{V}_j$ leads to improvements if this difference is negative for all $v\in\mathcal{V}_i$.

We proceed by a similar process to identify best pairs of nodes in $(\mathcal{V}_i, \mathcal{V}_j)$ to swap by trying all possible swaps, and updating $\mathcal{P}$ accordingly. Pair $(v_i^*,v_j^*) \in \mathcal{V}_i\times\mathcal{V}_j$ is \textit{the best pair to swap} if it maximizes 
\begin{multline*}
    \max(c(\mathcal{V}_i),c(\mathcal{V}_j))\\
    -\max(c(\mathcal{V}_i\backslash\{v_i\}\cup\{v_j\}),c(\mathcal{V}_j\backslash\{v_j\}\cup\{v_i\})).
\end{multline*}
 If this difference is negative for all $v_i \in \mathcal{V}_i$ and all $v_j \in \mathcal{V}_j$, then no swap between $\mathcal{V}_i$ and $\mathcal{V}_j$ leads to improvements.

To ensure that all pairs of subsets $(\mathcal{V}_i, \mathcal{V}_j)$ are checked for swaps and transfers, all pairs are initially marked to be checked for potential transfers or swaps between them. The algorithm then proceeds through all pairs in increasing order of indices, and a pair is unmarked if no transfer or swap reduces the maximum cost of these subsets. If a transfer or swap of nodes is found between two subsets $\mathcal{V}_i$ or $\mathcal{V}_j$ we then re-mark all pairs that contain either subset since more transfers-and-swaps may now be found. The process continues as long as there are marked pairs. While the original application of transfers-and-swaps in \cite{transfers-swaps} proves that the algorithm will eventually reach a state in which all pairs are unmarked, there is no guarantee that it will do so in the proposed adaptation. Thus, to prevent an infinite loop, we store all partitions that have already been tried; if the algorithm reaches a partition that has already been tried before, it returns this partition as the output. The above description is presented by the following algorithm.

\begin{algo}
    $\underline{\textsc{ \textbf{Algorithm 1: }transfers\_and\_swaps}(\mathcal{P}):}$\+
\\      Mark all pairs of subsets to be checked
\\      \textbf{while} there are marked pairs of subsets
\\      \textbf{and} $\mathcal{P}$ has not been tried before:\+
\\          \textbf{for} $\mathcal{V}_i, \mathcal{V}_j$ in marked transfers:\+
\\              $v^* \gets$ Best transfer from $\mathcal{V}_i$ to $\mathcal{V}_j$
\\              \textbf{if} $v^*$ exists:\+
\\                  Update $\mathcal{P}$: transfer $v^*$ from $\mathcal{V}_i$ to $\mathcal{V}_j$\-\-
\\          \textbf{for} $\mathcal{V}_i, \mathcal{V}_j$ in marked swaps:\+
\\              $(v_i^*,v_j^*) \gets$ Best swap between $\mathcal{V}_i$ and $\mathcal{V}_j$
\\              \textbf{if} $(v_i^*,v_j^*)$ exists:\+
\\                  Update $\mathcal{P}$: Transfer $v_i^*$ from $\mathcal{V}_i$ to $\mathcal{V}_j$ \\ 
and transfer $v_j^*$ from $\mathcal{V}_j$ to $\mathcal{V}_i$\-\-
\\          re-mark and unmark pairs to be checked\-
\\      \textbf{return} $\mathcal{P}$
\end{algo}

Each update of $\mathcal{P}$ in Algorithm 1 is made only if it reduces the maximum cost of two subsets. When two subsets are of similar cost, transferring a node whose existence increases the cost of a subset by a substantial amount to some other subset may not reduce this maximum. However, moving such \textit{high-cost nodes} may be of use in optimizing our partition. We describe our method for identifying such nodes and present it as Algorithm 2.

Consider a node $v$ in a subset $\mathcal{V}_i \in \mathcal{P}$. Let $\phi_i$ be the heuristic assignment on $\mathcal{V}_i$ and $\phi_i'$ be the heuristic assignment on $\mathcal{V}_i \setminus \{v\}$. We define \textit{the contribution} of a node $v \in \mathcal{V}_i$ as

\begin{equation}\label{contribution}
    \textsc{contribution}_{\mathcal{V}_i}(v) = \frac{\textsc{wlp}(\phi_i) - \textsc{wlp}(\phi_i')}{\textsc{wlp}(\phi_i)}.
\end{equation}

If the value calculated by \eqref{contribution} is greater than a certain fixed threshold $\alpha$, we call $v$ an \textit{outlier}, indicating that the existence of $v$ significantly increases the cost of $\mathcal{V}_i$. While appropriate tuning of $\alpha$ depends on the graph and the heuristics, we found that an $\alpha$ value of $0.13$ worked well for both the \textsc{greedy} and \textsc{nearest\_neighbor} heuristics. Once an outlier $v$ is identified, it is moved to subset $\mathcal{V}_j$ such that
$\mathcal{V}_j = \argmin_{\mathcal{V}_k \in \mathcal{P}} c\left( \mathcal{V}_k \cup \{v\} \right)$. Algorithm 2 thus takes in a partition $\mathcal{P}$ and some threshold $\alpha$ and returns an updated partition with outliers moved to new subsets.

\begin{algo}
    $\underline{\textsc{ \textbf{Algorithm 2: }transfer\_outliers}(\mathcal{P}, \alpha):}$\+
\\      \textbf{for} $i \gets 1..m$:\+
\\          \textbf{for} $v$ in $\mathcal{V}_i \setminus \{s\}$:\+
\\              \textbf{if} $\textsc{contribution}_{\mathcal{V}_i}(v) > \alpha$:\+
\\                  $\mathcal{V}_j \gets \argmin_{\mathcal{V}_k \in \mathcal{P}} c\left( \mathcal{V}_k \cup \{v\} \right)$ 
\\                  \textbf{if} $\mathcal{V}_i \neq \mathcal{V}_j$ :\+
\\                      Update $\mathcal{P}$: $\mathcal{V}_i \gets \mathcal{V}_i \setminus \{v\}, \mathcal{V}_j \gets \mathcal{V}_j \cup \{v\}$\-\-\-\-
\\      \textbf{return} $\mathcal{P}$
\end{algo}

We generate a random initial partition $\mathcal{P}$ given a graph $G$ and some number of agents $m$. Algorithm 3 then  iteratively alternates between Algorithm 1 and Algorithm 2 in order to find a local minimum over the space of possible partitions.

\begin{algo}
    $\underline{\textsc{ \textbf{Algorithm 3: }transfers\_swaps\_outliers}(G, m, \alpha):}$\+
\\      $\mathcal{P} \gets$ random starting partition of nodes $V$ in $G$ of $m$ agents
\\      $\mathcal{P}' \gets \textsc{transfers\_and\_swaps}\left( \mathcal{P} \right)$
\\      $\pi, \pi' \gets $ assignments based on $\mathcal{P}, \mathcal{P}'$
\\      \textbf{while} $\textsc{wlp\_sum}\left( \pi' \right) < \textsc{wlp\_sum}\left( \pi \right):$\+
\\          $\mathcal{P}' \gets \textsc{transfer\_outliers}\left( \mathcal{P}', \alpha \right)$
\\          $\mathcal{P}' \gets \textsc{transfers\_and\_swaps}\left( \mathcal{P}' \right)$
\\          $\pi, \pi', \gets $ assignments based on $\mathcal{P}, \mathcal{P'}$
\\          \textbf{if} $\textsc{wlp\_sum}\left( \pi' \right) < \textsc{wlp\_sum}\left( \pi \right):$\+
\\              $\mathcal{P} \gets \mathcal{P'}$\-\-
\\      $\pi \gets $ assignment based on $\mathcal{P}$ 
\\      \textbf{return} $\pi$
\end{algo}

We now move to validate the proposed algorithm on a set of random graphs, comparing it to benchmarks from \cite{estimation,prediction}, as well as on graphs inspired by a real-world storm scenario.

\section{Experimental Results}
\label{res}

In this section, we show that the proposed method consistently beats prior strategies in finding solutions to \eqref{optimization-sum}. However we emphasize that neither our work, nor \cite{estimation, prediction}, have any formal upper bounds on the optimality of solutions. The worst case runtime of our algorithm is non-polynomial with respect to $n$. In practice, we found that with even a personal laptop computer or freely available server instances such as AWS (single thread vCPU), it is feasible to run Algorithm 3 on graphs of hundreds of nodes with dozens of agents. Namely the the randomized environments in Section \ref{rangen} each took under 5 minutes to compute.

We implemented the proposed algorithms in Python 3. The code and other materials needed to run the simulations are available at \url{https://github.com/leadcatlab/MWLP-Storm-Repair}.

In order to describe the comparative strength of our strategy, we begin by briefly recounting strategies in prior work \cite{estimation, prediction}.

\subsection{Benchmark Strategies}
\label{pristr}

Both \cite{estimation} and \cite{prediction} consider assignments of targets for repair crews in order to minimize \eqref{wlp_sum}. However, because their focus is on assignment analysis for large-scale graphs rather than the synthesis of optimal assignments, the strategies they consider are simple and based on human intuition.

A strategy used in \cite{estimation} is a \textit{nearest neighbor search}, which we denote by \textsc{nearest\_neighbor\_assignment}. At each step, each agent will be assigned to the nearest node that has not already been visited by another agent.

The second strategy, introduced in \cite{prediction}, is a purely \textit{greedy} approach, which we denote by \textsc{greedy\_assignment}. At each step, each agent will visit the node of highest importance that has not already been visited to another agent.

We note the similarities and differences between \textsc{greedy\_assignment} and \textsc{nearest\_neighbor\_assignment} as defined above and our heuristics \textsc{greedy} and \textsc{nearest\_neighbor} as used in our algorithm. While the underlying idea for them is the same, the latter pair of heuristics operate on each partition element separately, whereas the former methods operate without explicitly partitioning their targets.

Paper \cite{prediction} also considers a mixed randomized approach. Namely, repair crews are a priori separated into two sets. Half of the agents follow \textsc{greedy\_assignment}, while the other half utilize a \textit{randomized neighbor search}. These agents travel to a random neighbor of their current target within a fixed \textit{search radius}; if no such neighbor is within this radius, the agents proceed to their nearest neighbor. We refer to this strategy as \textsc{greedy\_random\_assignment}. While \cite{prediction} experiments with multiple choices for the fixed search distance for the second set of crews, we found that a radius of $d/4$, where $d$ is the difference between the longest and shortest edge in the current graph, provided some of the best results.

We now proceed to simulate the performance of our assignment algorithm and compare it to the described benchmarks.

\subsection{Performance on Random Environments}
\label{rangen}

We compare the proposed transfers-and-swaps based Algorithm 3, using either the $\textsc{greedy}$ or $\textsc{nearest\_neighbor}$ heuristics, to the strategies from prior work described in Section \ref{pristr}. We consider their performance over 100 environments modeled by randomly generated graphs with the following parameters evocative of a small storm in an urban area: $m=20$, indicating 20 repair crews, $n=201$, indicating $200$ repair locations and one joint starting location, and $w(v)\in [1,1500]$ for all $v\in V$, indicating the population that depends on service at location $v$. The edge-weights $d(e)$ are constructed by adding the travel times between any two locations --- chosen to be between 30 and 60 minutes --- to the repair times at the destination node. The repair times are chosen at random while constructing the graph, as modeled by Table~\ref{repair_times} introduced in \cite{prediction}.

\begin{table}[H]
\caption{Repair times from \cite{prediction}}
\label{repair_times}
\centering
\begin{tabular}{llr}
\hline
Minimum $w(v)$ & Maximum $w(v)$ & Repair Time Range (hours) \\ \hline
0              & 10             & 2-4                       \\
10             & 100            & 2-6                       \\
100            & 1000           & 3-8                       \\
1000           &                & 5-10                      \\ \hline
\end{tabular}
\end{table}

Table~\ref{abbrev} introduces the abbreviations used for the five algorithms in subsequent figures.
\begin{table}[H]
\caption{Labels for algorithms for use in Fig.~\ref{wait_time_plot}--\ref{champaign_unvisited}}
\label{abbrev}
\centering
\begin{tabular}{lr}
\textsc{greedy\_assignment}                                          & GA   \\
\textsc{nearest\_neighbor\_assignment}                               & NNA  \\
\textsc{greedy\_random\_assignment}                                  & GRA  \\
\textsc{transfers\_swaps\_outliers} with \textsc{greedy}             & TSG  \\
\textsc{transfers\_swaps\_outliers} with \textsc{nearest\_neighbor}  & TSNN
\end{tabular}
\end{table}

\begin{figure}[H]
    \centering
    \includegraphics[scale=0.33]{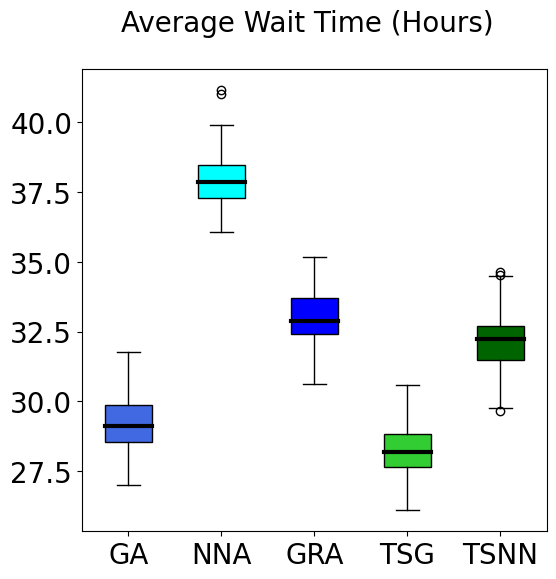}
    \caption{Box plot of average wait times in hours for all 100 environments across all five algorithms. The middle black line represents the median, the range of the colored box represents the interquartile range (IQR), the length of whiskers is 1.5 IQR. Any circles beyond the whiskers are outlying values.}
    \label{wait_time_plot}
\end{figure}

The primary goal of the proposed algorithm is to minimize the total weighted latency incurred by an agent assignment, defined as in \eqref{wlp_sum}. Recall that \eqref{wlp_sum} is a weighted sum of waiting times for every location. Given that in this case the weights indicate population counts, \eqref{wlp_sum} induces a natural notion of an \textit{average wait time}. We define such a time for an assignment $\pi = \set{\phi_1, \ldots, \phi_m}$ by 
\begin{equation}\label{wait_time}
    \textsc{wait}(\pi) = \frac{\textsc{wlp\_sum}(\pi)}{\sum_{i = 1}^n w(v_i)}.
\end{equation} 
The value of $\textsc{wait}(\pi)$ represents the average time a customer waits for their service to be restored, given assignment $\pi$. Statistical analysis of the average wait times over all 100 environments is given in Fig.~\ref{wait_time_plot}. Naturally, the statistics of the value of \eqref{wlp_sum} are qualitatively the same.

From Fig.~\ref{wait_time_plot}, we see that our proposed algorithm performs better than prior strategies. Namely, the proposed transfers-and-swaps algorithm with the \textsc{greedy} heuristic (TSG) performs the best out of all 5 algorithms. In general, heuristics based on importance (GA, TSG) perform better than those based on distance (NNA, TSNN). We also see that TSNN performs significantly better than NNA.

\begin{figure}[H]
    \centering
    \includegraphics[scale=0.33]{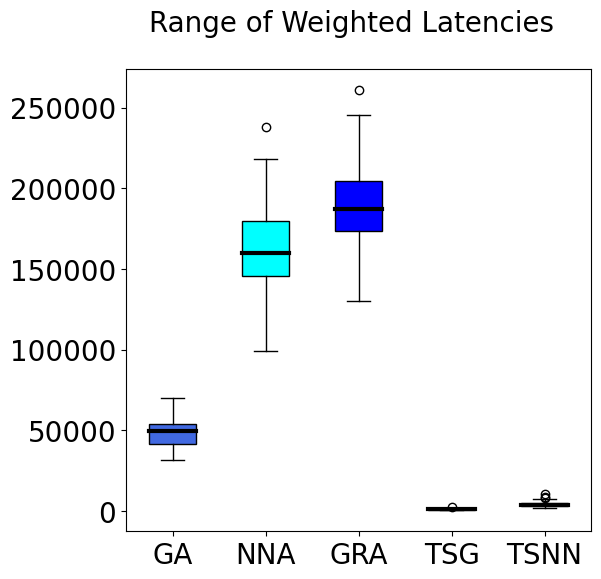}
    \caption{Box plot of ranges of weighted latencies for all 100 environments across the considered algorithms. Elements of the box plots are the same as in Fig.~\ref{wait_time_plot}.}
    \label{range_plot}
\end{figure}

Fig.~\ref{range_plot} shows that, along with providing the best average performance, the proposed transfers-and-swaps approach divides the work among different crews more evenly. To describe this feature, for a given assignment $\pi = \set{\phi_1, \ldots, \phi_m}$ we define
\begin{equation*}\label{range}
    \textsc{range}(\pi) = \max_{\phi_i \in \pi} \left\{ \textsc{wlp}(\phi_i) \right\} - \min_{\phi_i \in \pi} \left\{ \textsc{wlp}(\phi_i) \right\},
\end{equation*}
thus measuring the difference between the crew with the highest weighted latency and the lowest weighted latency. Our algorithms yield significantly lower \textsc{range} than the benchmarks, indicating that cost is distributed more evenly.

After these promising results on a large number of fully synthetic environments, we now proceed to consider an environment built from real-world data.

\subsection{Case Study: Champaign}

We evaluate our algorithm by comparing it to existing benchmarks on an environment modeling a significant weather event in Champaign, Illinois, USA. We use data on the road network and travel times obtained from OpenStreetMap, assuming that all crews travel between repair locations by road at a constant speed of 25 miles per hour ($\approx$ 40 kilometers per hour). This assumption is in line with the model of \cite{prediction}.

We generate 25 target sets chosen at random from the road network data, with the same parameters as in Section~\ref{rangen}: $m=20$ and $n=201$. The targets, as well as the starting state, are chosen at random from intersections in populated areas of the city. Their importance levels correspond to the populations of adjacent population tracts, obtained from the US Census Bureau database. Finally, the edge-weights are travel times between the nodes, obtained by computing shortest paths between corresponding intersections on the road network.

The area of Champaign is much smaller than the areas studied in \cite{prediction, estimation}. The travel times between targets are minuscule compared to the repair times in Table~\ref{repair_times}; given the size of Champaign, they range between 1 and 30 minutes. Since repair times of individual targets do not depend on an assignment and in that sense ``dilute'' the comparison between assignment algorithms, we describe the performance of our proposed method by setting all repair times to 0.

From Section~\ref{rangen}, we conclude that the \textsc{greedy\_assignment} (GA) benchmark and the proposed  \textsc{transfers\_swaps\_outliers} with \textsc{greedy} (TSG) may yield the best assignments. Thus, in this section we focus on comparing the performance of these two algorithms.

\begin{figure}[H]
    \centering
    \begin{subfigure}{.25\textwidth}
        \centering
        \includegraphics[scale=0.33]{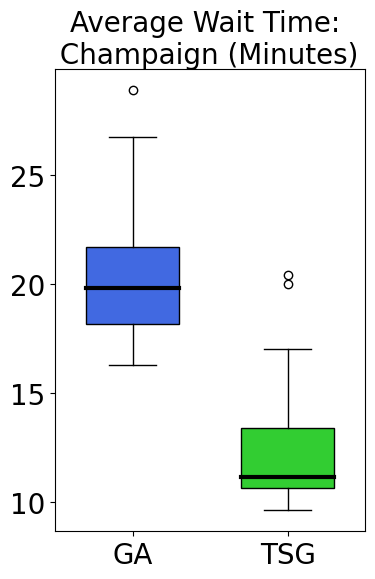}
    \end{subfigure}%
    \begin{subfigure}{.25\textwidth}
        \centering
        \includegraphics[scale=0.33]{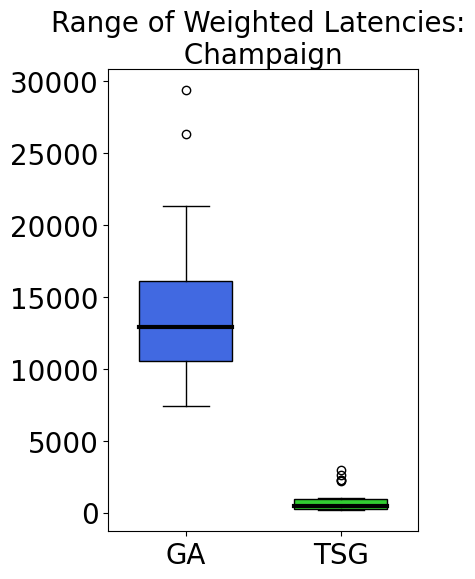}
    \end{subfigure}
    \caption{Box plots of average wait times and ranges of weighted latencies for all 25 Champaign target sets for the benchmark greedy algorithm (GA) and proposed TSG algorithm. Elements of the box plot are the same as in Fig.~\ref{wait_time_plot}.}
    \label{combination plot}
\end{figure}

Fig.~\ref{combination plot} compares the success of crew assignments using the proposed algorithm to the benchmark strategies, using the same metrics as in Section~\ref{rangen}. We again remark that plotting the values of \eqref{wlp_sum} produces a virtually identical graph to that of average wait times, with different scales.

The performance of these algorithms over the urban environment aligns with that of the previous numerical example: not only does TSG significantly outperform the best benchmark algorithm, but the difference in average wait time is even more prominent in the case when repair times are small compared to travel times. The plot of ranges shows that again the proposed algorithm provides a more even distribution of work among crews compared to the benchmark.

\begin{figure}[H]
    \centering
    \vspace{-30pt}
    \includegraphics[scale=0.33]{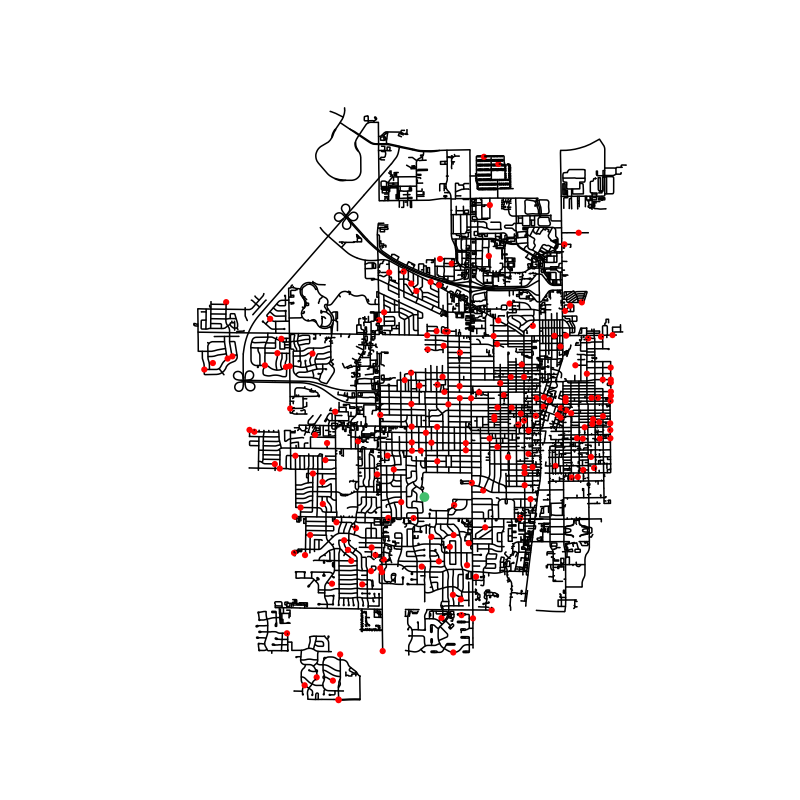}
    \vspace{-20pt}
    \caption{Plot of a random set of targets in Champaign. Red dots signify targets and the green dot signifies the repair depot.}
    \label{champaign_targets}
\end{figure}

\begin{figure}[t]
    \centering
    \includegraphics[scale=0.33]{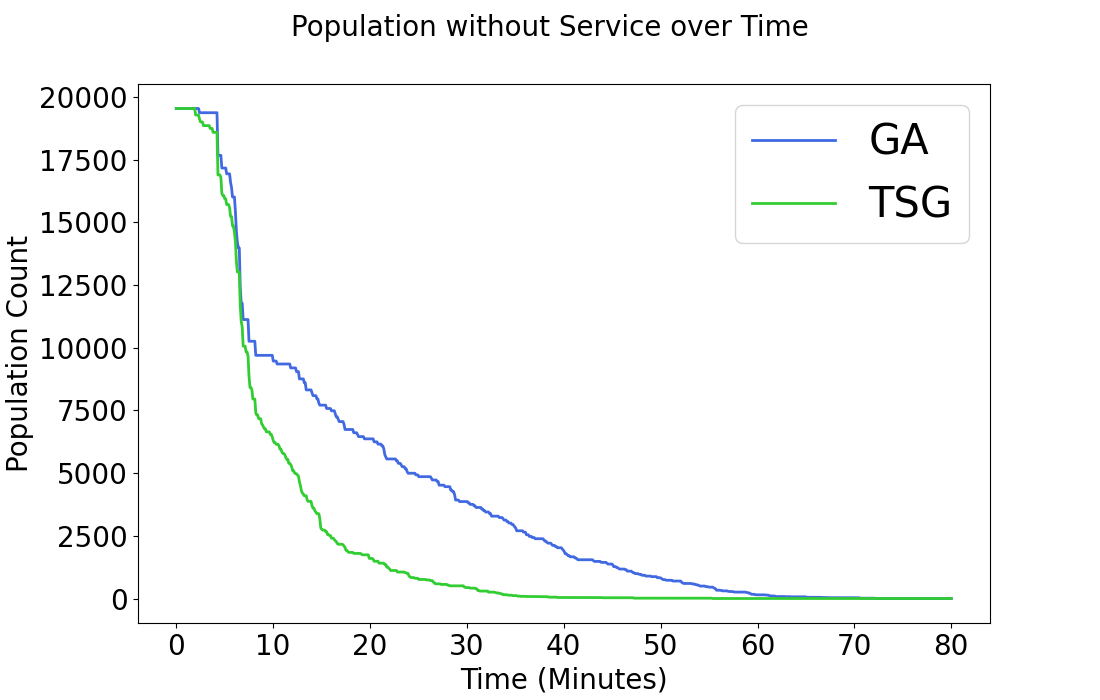}
    \caption{Line plot of population lacking service over time for one set of target locations in Champaign.}
    \label{champaign_unvisited}
\end{figure}

Finally, we focus on a particular set of targets shown in Fig.~\ref{champaign_targets}. A standard \cite{prediction, estimation} description of assignment performance is to examine the population that still lacks service as a function of time. We provide such a plot in Fig.~\ref{champaign_unvisited}.

The proposed algorithm indeed results in customers receiving service much faster than the benchmark, and completes the overall repair task $\approx$ \textit{50\% quicker}.

\section{Conclusions and Future Work}
\label{future}

This paper deals with the problem of optimal repair crew assignment to repair locations in an post-disaster repair scenario. By using an agent-based model (ABM), we interpret this problem as a novel Minimum Weighted Latency Problem for $m$ agents ($m$-MWLP). To solve it, we propose an approximation heuristic that a priori determines agent assignments based on a partition of target locations which is incrementally improved using a transfers-and-swaps method. The proposed heuristic vastly outperforms the policies used in prior work.

 
Despite initial success, this work is preliminary in many ways. Future work should remove the assumptions mentioned in Section~\ref{assumptions} such as travel time determinism or complete prior knowledge of all system data. A possible way to do so is through the framework of partially observable graphs --- for unknown edge-weights or node-weights --- and semi-Markov decision processes, for stochastic travel times.

In addition to the limitations posed by current assumptions, there is space for improving the existing algorithm. Namely, we remark that the transfers-and-swaps framework allows for the use of arbitrary heuristics for partition cost evaluation. While we empirically showed that \textsc{greedy} and \textsc{nearest\_neighbor} heuristics already outperform prior work, there may exist a more involved, yet computationally feasible, cost function that better approximates the optimal cost of a partition. A possible path forward would be a heuristic that takes into account both target importance and distance, unlike two extreme options of \textsc{greedy} and \textsc{nearest\_neighbor}.

Finally, even though the preliminary empirical results are promising, current work lacks theoretical grounding on \textit{optimality} and \textit{scalability}. There are currently no upper bounds on the quality of the approximations using the proposed heuristics. Additionally, the proposed algorithm has a non-polynomial run-time. Running the algorithm is feasible for a few dozen crews and several hundreds of targets. However, the large-scale storms analyzed in \cite{prediction, estimation} are not practically analyzed. One way to greatly reduce run-time would be to have a method to disqualify some amount of partitions at each step, in a way similar to that of a branch-and-bound method would. Doing so would lend itself to improved scalability since the majority of the computational complexity arises from the number of possible partitions of a graph, and thus is a natural direction for future work. 

\section*{Acknowledgments}

We thank Pranay Thangeda for his help in identifying bugs in early versions of the code for this paper as well as providing baseline code for the Champaign graph. We also thank Diego Cerrai for clarifying previous work on this topic and kindly sharing his recent results.

\bibliography{refs}

\begin{thebibliography}{10}
\providecommand{\url}[1]{#1}
\csname url@samestyle\endcsname
\providecommand{\newblock}{\relax}
\providecommand{\bibinfo}[2]{#2}
\providecommand{\BIBentrySTDinterwordspacing}{\spaceskip=0pt\relax}
\providecommand{\BIBentryALTinterwordstretchfactor}{4}
\providecommand{\BIBentryALTinterwordspacing}{\spaceskip=\fontdimen2\font plus
\BIBentryALTinterwordstretchfactor\fontdimen3\font minus
  \fontdimen4\font\relax}
\providecommand{\BIBforeignlanguage}[2]{{%
\expandafter\ifx\csname l@#1\endcsname\relax
\typeout{** WARNING: IEEEtran.bst: No hyphenation pattern has been}%
\typeout{** loaded for the language `#1'. Using the pattern for}%
\typeout{** the default language instead.}%
\else
\language=\csname l@#1\endcsname
\fi
#2}}
\providecommand{\BIBdecl}{\relax}
\BIBdecl

\bibitem{extreme-events-happen}
A.~Castillo, ``Risk analysis and management in power outage and restoration: A
  literature survey,'' \emph{Electric Power Systems Research}, vol. 107, pp.
  9--15, 2014.

\bibitem{climate-change}
S.~Ćurčić, C.~Özveren, L.~Crowe, and P.~Lo,
  ``\BIBforeignlanguage{English}{Electric power distribution network
  restoration: a survey of papers and a review of the restoration problem},''
  \emph{\BIBforeignlanguage{English}{Electric Power Systems Research}},
  vol.~35, no.~2, pp. 73--86, 1995.

\bibitem{estimation}
T.~Walsh, T.~Layton, D.~Wanik, and J.~Mellor, ``Agent based model to estimate
  time to restoration of storm-induced power outages,'' \emph{Infrastructures},
  vol.~3, no.~3, 2018.

\bibitem{linear-program-positions}
S.~D. Whipple, ``Predictive storm damage modeling and optimizing crew response
  to improve storm response operations,'' Master's thesis, Massachusetts
  Institute of Technology, 2014.

\bibitem{linear-program-networks}
Y.~Ge, L.~Du, and H.~Ye, ``Co-optimization approach to post-storm recovery for
  interdependent power and transportation systems,'' \emph{Journal of Modern
  Power Systems and Clean Energy}, vol.~7, no.~4, pp. 688--695, 2019.

\bibitem{linear-program-stations}
D.~Chang, D.~Shelar, and S.~Amin, ``{DER} allocation and line repair scheduling
  for storm-induced failures in distribution networks,'' in \emph{IEEE
  International Conference on Communications, Control, and Computing
  Technologies for Smart Grids}, 2018.

\bibitem{prediction}
T.~C. Walsh, A.~Spaulding, and D.~Cerrai, ``Predicting outage restoration in
  advance of storms impact,'' 2022, under review.

\bibitem{transfers-swaps}
I.~Vandermeulen, R.~Gro\ss{}, and A.~Kolling, ``Balanced task allocation by
  partitioning the multiple traveling salesperson problem,'' in \emph{18th
  International Conference on Autonomous Agents and Multiagent Systems}, 2019,
  pp. 1479--1487.

\bibitem{MWLP-cases}
B.~Y. Wu, ``Polynomial time algorithms for some minimum latency problems,''
  \emph{Information Processing Letters}, vol.~75, no.~5, pp. 225--229, 2000.

\bibitem{ride-and-delivery}
A.~Das, S.~Gollapudi, A.~Kim, D.~Panigrahi, and C.~Swamy, ``Minimizing latency
  in online ride and delivery services,'' in \emph{27th International World
  Wide Web Conference}, 2018, pp. 379--388.

\bibitem{MLP-DP}
B.~Y. Wu, Z.-N. Huang, and F.-J. Zhan, ``Exact algorithms for the minimum
  latency problem,'' \emph{Information Processing Letters}, vol.~92, no.~6, pp.
  303--309, 2004.

\bibitem{the-MLP}
A.~Blum, P.~Chalasani, D.~Coppersmith, B.~Pulleyblank, P.~Raghavan, and
  M.~Sudan, ``The minimum latency problem,'' in \emph{26th Annual ACM Symposium
  on Theory of Computing}, 1994, pp. 163--171.

\bibitem{MLP-NP-Complete}
S.~Sahni and T.~Gonzalez, ``P-complete approximation problems,'' \emph{Journal
  of the ACM}, vol.~23, no.~3, pp. 555–--565, 1976.

\bibitem{fixed-graph}
E.~Koutsoupias, C.~Papadimitriou, and M.~Yannakakis, ``Searching a fixed
  graph,'' in \emph{23rd International Colloquium on Automata, Languages, and
  Programming}, 1996, pp. 280--289.

\bibitem{delivery-tree}
E.~Minieka, ``The delivery man problem on a tree network,'' \emph{Annals of
  Operations Research}, vol.~18, no.~1, pp. 261--266, 1989.

\bibitem{TRP-Complexity}
A.~Garc\'{i}a, P.~Jodr\'{a}, and J.~Tejel, ``A note on the traveling repairman
  problem,'' \emph{Networks}, vol.~40, no.~1, pp. 27--31, 2002.

\bibitem{approx-and-random}
J.~Bossek, K.~Casel, P.~Kerschke, and F.~Neumann, ``The node weight dependent
  traveling salesperson problem: Approximation algorithms and randomized search
  heuristics,'' in \emph{Genetic and Evolutionary Computation Conference},
  2020, p. 1286–1294.

\bibitem{parallel-programming}
Z.~Wei, ``New methods for solving the minimum weighted latency problem,'' Ph.D.
  dissertation, University of Alberta, 2018.

\bibitem{paths-and-trees}
K.~Chaudhuri, B.~Godfrey, S.~Rao, and K.~Talwar, ``Paths, trees, and minimum
  latency tours,'' in \emph{44th Annual IEEE Symposium on Foundations of
  Computer Science}, 2003, pp. 36--45.

\bibitem{transfers-swaps-MDP}
F.~Nawaz and M.~Ornik, ``Multi-agent, multi-target path planning in {Markov}
  decision processes,'' \emph{arXiv preprint arXiv:2205.15841 [math.OC]}, 2022.

\end{thebibliography}
\bibliographystyle{IEEEtran}

\end{document}